\documentclass[twocolumn,nobibnotes,showkeys,nofootinbib,superscriptaddress,amsmath,amssymb]{revtex4}

\usepackage{graphicx}
\usepackage{subfigure}
\usepackage{bm}
\newcommand{\tdm}[1]{\mbox{\boldmath$#1$}}

\def\be{\begin{equation}}
\def\ee{\end{equation}}
\def\bea{\begin{eqnarray}}
\def\eea{\end{eqnarray}}
\def\beeq{\begin{eqnarray}}
\def\eeeq{\end{eqnarray}}
\newcommand{\eq}{&=&}

\begin{document}
\title{\bf $F_L$ proton structure function from the unified DGLAP/BFKL approach}

 \author{K. Golec-Biernat}
  \email{golec@ifj.edu.pl}
   \affiliation{Institute of Nuclear Physics PAN, Krak\'ow, Poland}
    \affiliation{Institute of Physics, Rzesz\'ow University, Rzesz\'ow, Poland}

 \author{A. M. Sta\'sto}
  \email{astasto@phys.psu.edu}
   \affiliation{Penn State University, Physics Department, University Park, PA 16802, USA}
    \affiliation{RIKEN Center, Brookhaven National Laboratory, Upton, NY 11973, USA}
     \affiliation{Institute of Nuclear Physics PAN, Krak\'ow, Poland}

\date{\today}

\begin{abstract}
We compute the longitudinal proton structure function $F_L$ from the $k_T$ factorization scheme, 
using the unified DGLAP/BFKL resummation approach at small $x$ for the unintegrated gluon density.
The differences between the $k_T$ factorization, collinear factorization and dipole approaches are analyzed and discussed.  The comparisons with the HERA data are made and  
predictions for the proposed Large Hadron-Electron Collider are also provided.
\end{abstract}



\maketitle

\section{Introduction}

The longitudinal nucleon structure function $F_L$, measured in the deep inelastic lepton-nucleon scattering, is proportional to the cross section for the interaction of the longitudinally polarized virtual photon  with a nucleon.
This observable is of particular interest since it is directly sensitive to the nucleon  gluon distribution.
In the naive quark-parton model $F_L$ vanishes (the Callan-Gross relation). This is due to  the quark spin $1/2$ and the fact that  the struck quark
has limited transverse momentum in the naive parton model. In the QCD improved parton model, however,  the gluon interactions cause the average  quark
transverse momentum    $\langle \kappa_T^2 \rangle$ to
grow with increasing value of the (minus) photon virtuality $Q^2$. As a result, $F_L$ acquires a nonzero leading twist contribution proportional to $\alpha_s(Q^2)$. 
At small values of the Bjorken variable $x$, $F_L$ is driven mainly by gluons through the transition $g \rightarrow q \bar q$.
Therefore, it can be used for the extraction  
of the gluon distribution in a nucleon providing a crucial test of the validity of perturbative QCD in this kinematical range.

The experimental  determination of $F_L$ is in general difficult  and requires a measurement of the
inelastic cross section at the same values of $x$ and $Q^2$ but for different center-of-mass energy of the incoming beams.
This was achieved at the DESY electron-proton collider HERA by changing the proton beam energy with the lepton beam energy fixed. The structure function $F_L$ was measured both by the H1 \cite{:2008tx} and ZEUS \cite{ZEUSFL} collaborations in the $Q^2$ range of $12-90$ and $24-110~{\rm GeV}^2$, respectively.

At small $x$,  the nucleon  structure functions receive large logarithmic corrections coming from  resummation of large powers of $\alpha_s \ln 1/x$. This  procedure goes beyond the standard collinear factorization and is achieved by the use of the $k_T$ factorization formalism \cite{Catani:1990xk,Collins:1991ty} with the unintegrated gluon density found as a solution to the Balitsky-Fadin-Kuraev-Lipatov (BFKL) \cite{Fadin:1975cb,Lipatov:1976zz,Balitsky:1978ic} or Ciafaloni-Catani-Fiorani-Marchesini (CCFM) evolution equations \cite{Ciafaloni:1987ur,Catani:1989yc,Catani:1989sg,Marchesini:1994wr}. Since the small $x$ expansion receives large corrections at higher orders,  resummation at small $x$ is in general necessary in order to obtain predictions which are in agreement with data.

The objective of this paper  is the calculation of $F_L$ within the $k_T$ factorization formalism using the unintegrated gluon density obtained from the Kwieci\'nski-Martin-Sta\'sto  (KMS) approach  \cite{Kwiecinski:1997ee}, which provides a convenient framework for the unification of the conventional
Dokshitzer-Gribov-Lipatov-Altarelli-Parisi (DGLAP)  
\cite{Gribov:1972ri,Altarelli:1977zs,Dokshitzer:1977sg} and small $x$ BFKL  evolution equations. From the point of view of the small $x$ hierarchy, the KMS approach includes important effects of higher order resummation.
In addition, we systematically analyze the relation between this approach and the collinear and dipole approaches, investigating the role of different contributions to $F_L$ in various kinematical regions.  We especially emphasize the role of the exact gluon kinematics in the 
$k_T$ factorization formulae and demonstrate numerically that this kinematics  have a sizable effects on the predictions for $F_L$, and thus, on the extracted gluon density. We compare our computations with the experimental  data at small $x$ from the H1 \cite{:2008tx} and ZEUS \cite{ZEUSFL} collaborations.

The paper is organized as follows. In Sec.~\ref{sec:flkt} we recall the $k_T$ factorization formalism  for the longitudinal  structure function. In Sec.~\ref{sec:unified} we review the unified BFKL/DGLAP approach for the unintegrated gluon density which includes important resummation effects at small $x$.  In Sec.~\ref{sec:colldipole} we discuss the relation of 
the $k_T$ factorization with the collinear and  dipole approaches. In Sec.~\ref{sec:numerics} we present a systematic numerical analysis of the various approaches and compare them with the HERA data, as well as provide the extrapolations to the LHeC (Large Hadron-Electron Collider \cite{Klein:2008zz}) energies. In the last section we summarize our conclusions.

\begin{figure*}[t]
\centerline{
\includegraphics[width=0.25\textwidth]{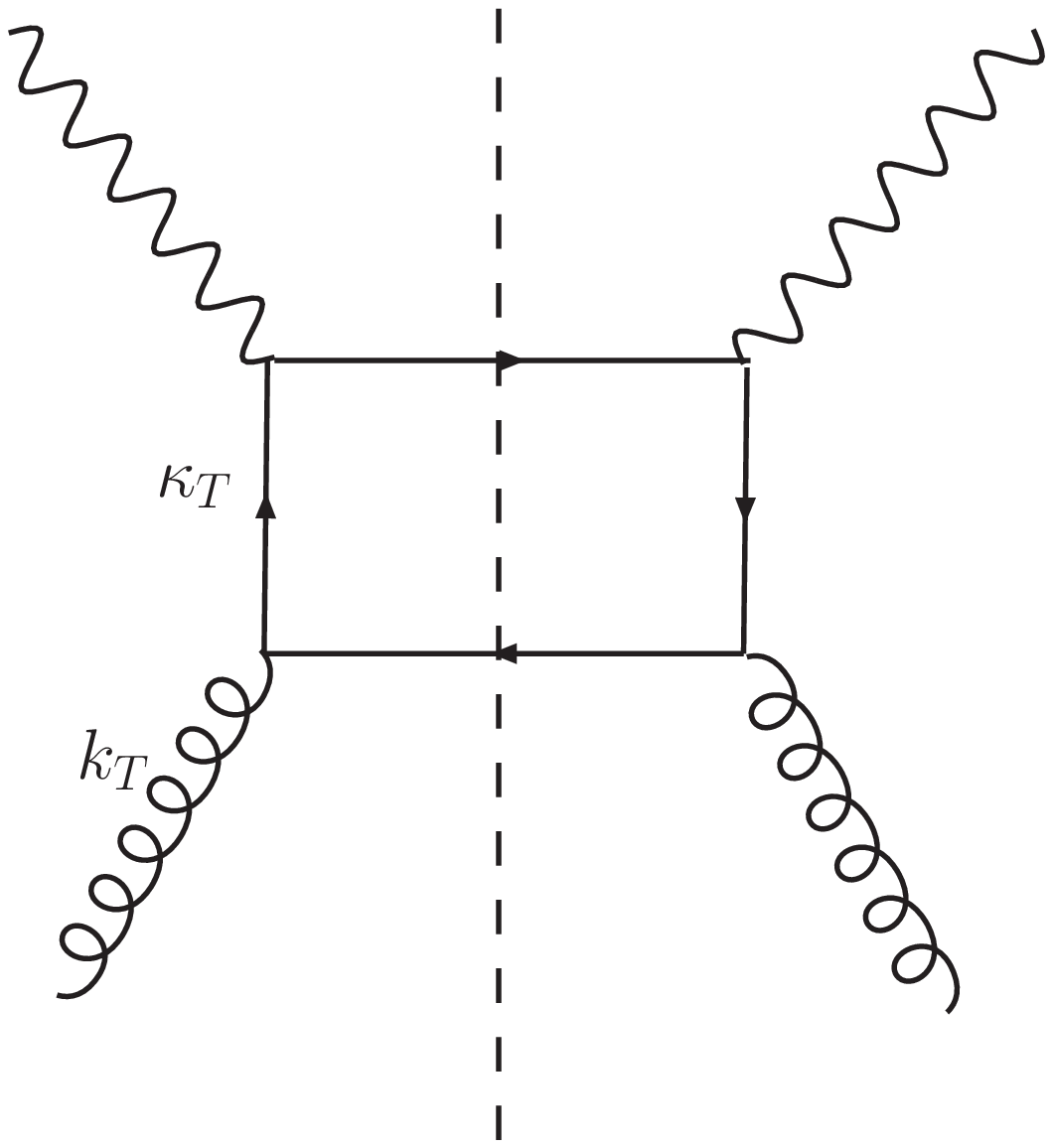}\hspace*{2cm}
\includegraphics[width=0.25\textwidth]{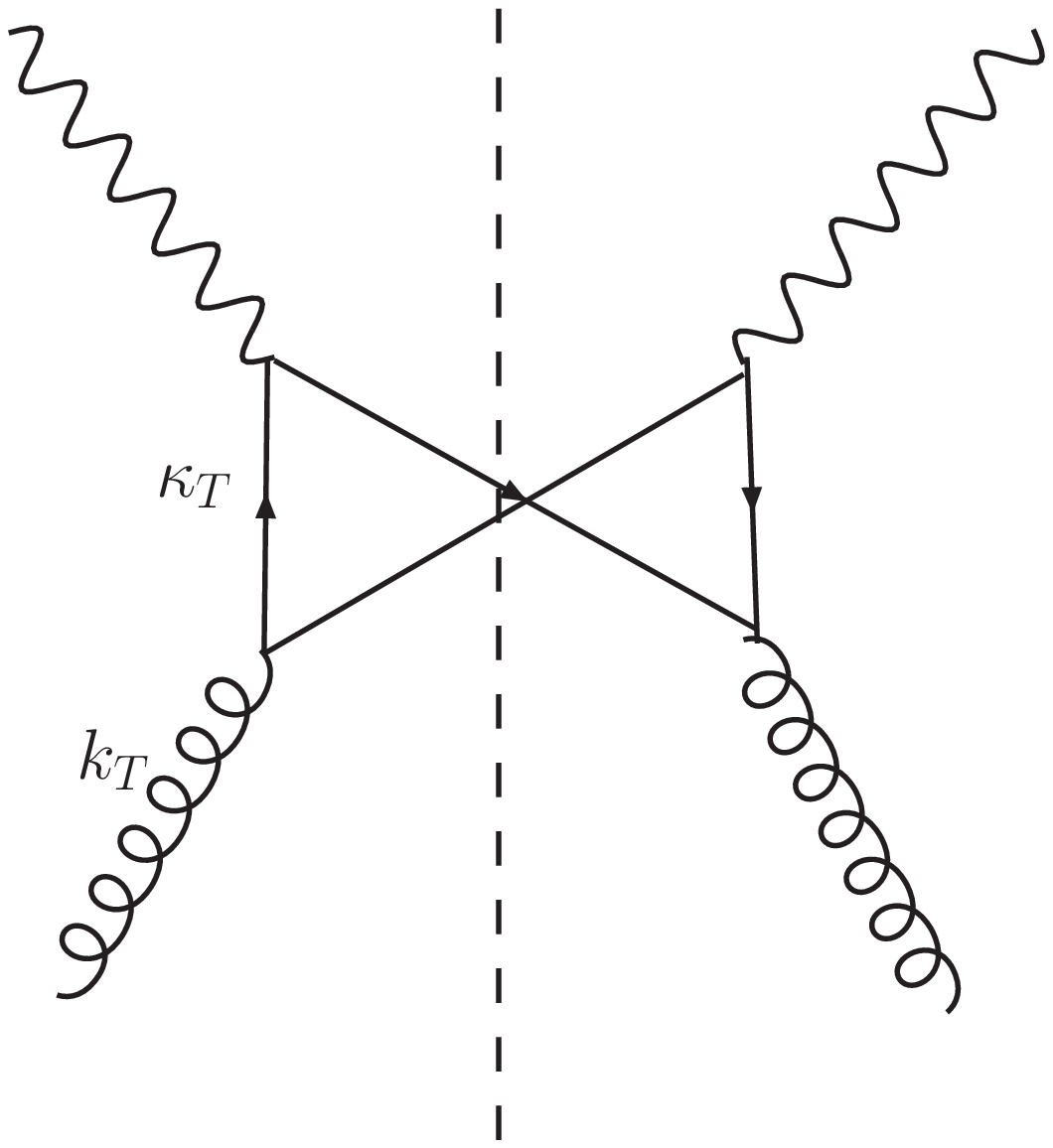}}
\caption{Quark box diagrams in the photon-gluon fusion process used in the  $k_T$-factorization formula.}
\label{fig:qbox}
\end{figure*}

\section{$F_L$ from the $\tdm{k_T}$ factorization approach.}
\label{sec:flkt}

In the limit of the high center-of-mass energy, or equivalently at small values of Bjorken  $x$, the nucleon structure functions can be computed from the $k_T$ factorization approach 
\cite{Catani:1990xk,Collins:1991ty}. 
Basic diagrams for the boson-gluon fusion which are taken into account in the high energy limit are depicted in Fig.~\ref{fig:qbox}. The gluon is off-shell with its virtuality dominated by the transverse momentum $k_T\equiv\tdm{k}$, see Fig~\ref{fig:qbox}.
Since we are interested in $F_L$, the photon is  longitudinally polarized. Thus, in the $k_T$ factorization approach the longitudinal  structure function is  then given by 
\begin{widetext}
\beeq
 F_L(x,Q^2)=2
{Q^4\over \pi^2} \sum _qe_q^2
\int {dk^2\over k^4} \int_0^1 d\beta \int d^2{\tdm{\kappa^{\prime}}} \,
\alpha_s(\mu^2) \, \beta^2\, (1-\beta)^2\,\frac{1}{2}
\left({1\over D_{1q}} - {1\over D_{2q}}\right)^2 f\left(\frac{x}{z},k^2\right)~~\;,
\label{flint}
\eeeq
\end{widetext}
where the denominators $D_{1q}$ and $D_{2q}$ read
\bea
\label{eq:q1}
D_{1q} & = & \kappa^2 \: + \: \beta (1 - \beta) Q^2 \: + \: m_q^2  \; ,
\label{eq:denominator1} \\
D_{2q} & = & (\mbox{\boldmath $\kappa$} - \mbox{\boldmath
$k$})^2 \: + \: \beta (1 - \beta) Q^2 \: + \: m_q^2 \; ,
\label{eq:denominator2} 
\eea
the quark transverse momentum  $\kappa_T\equiv\tdm{\kappa}$, 
the shifted transverse momentum is given by $\tdm{\kappa^{\prime}}=\tdm{\kappa} - (1 - \beta) \tdm{k}$,
and the argument of the unintegrated gluon density $f({x}/{z},k^2)$ is defined to be
\be
\frac{x}{z}\equiv x_g  \equiv x \left( 1 \: + \: \frac{\kappa^{\prime 2} +
m_q^2}{\beta (1 - \beta) Q^2} \: + \: \frac{k^2}{Q^2} \right ) .
\label{eq:xgluon}
\ee
The variable $\beta$ is the corresponding Sudakov
parameter appearing in the quark momentum decomposition, 
\beeq
\kappa&=&x_q\,p^\prime-\beta\, q^\prime+\tdm{\kappa} \; 
\\
x_q&=&x\left(1+{m_q^2+\tdm{\kappa}^2 \over (1-\beta)Q^2}\right) ,
\label{decomp}
\eeeq
with the following light-like base vectors
\be
p^\prime=p-{M^2 x \over Q^2}q\,,~~~~~~~~~~~~~q^\prime=q+xp \; 
\ee
where $M$ denotes the nucleon mass, $m_q$ is the quark mass which we keep nonzero only for charm
quark,  $p$ is the target-proton four-momentum and $q$ is the virtual photon four-momentum.
The  argument in the strong coupling constant is taken to be  
\be
\mu^2 = \kappa^{\prime2}+k^2+m_q^2\,.
\ee
The integration over the gluon virtuality $k^2$ in eq.~(\ref{flint}) needs special care in the 
low momenta region, $k^2<k_0^2\simeq 1~{\rm GeV}^2$. We will discuss this important element of our presentation in the forthcoming sections.

The function $f(y,k^2)$ is the  unintegrated gluon distribution, which
in the small $x$ limit is related to the conventional (integrated) gluon distribution 
$g(y,\mu^2)$ by
\begin{equation} 
yg(y,\mu^2)=\int^{\mu^2}\frac{dk^2}{k^2}\, f(y,k^2)\;.
\label{intg}
\end{equation} 
The integration limits in (\ref{flint}) are  constrained by the  condition $x_g<1$ while
the condition $x_g>x$ is automatically satisfied from eq.~(\ref{eq:xgluon}).
We note that in the strict high-energy limit the argument
of the unintegrated gluon distribution would be set to the Bjorken $x$. This is also the usual procedure in the dipole picture approach which we discuss in Sec.~\ref{sec:colldipole}.
Here, we take into account the effects of exact kinematics
which results in  the shift of the gluon  $x_g$ to larger values than $x$. This is related to the fact that the energy needed to produce the  $q\bar{q}$ pair  is non-negligible even when the total center-of-mass energy is very large. Although this effect is non-leading in the leading logarithmic small $x$ approximation, it is nevertheless numerically quite important, as we will illustrate 
 in Sec.~\ref{sec:numerics}. 

It also  has been shown in the dipole picture that by including the exact kinematics in the argument of the gluon distribution,  the transverse size of the quark-antiquark dipole is no longer conserved \cite{Bialas:2000xs,Bialas:2001ks}, see Sec.~\ref{sec:colldipole:b}.

\section{Unified DGLAP and BFKL equations}
\label{sec:unified}

The main  input to the $k_T$ factorization formula is the unintegrated gluon distribution $f(y,k^2)$. At small $x$, this distribution can  be found from the solution to the BFKL \cite{Fadin:1975cb,Lipatov:1976zz,Balitsky:1978ic} or the CCFM equations \cite{Ciafaloni:1987ur,Catani:1989yc,Catani:1989sg,Marchesini:1994wr}. 
These equations give  predictions for the unintegrated gluon density as a function of the transverse momentum squared $k^2$ and $x$ (and also an external scale $Q$ in the case of the  CCFM equation) provided $x\ll 1$. A more rigorous approach which includes the operator definitions of  unintegrated gluon densities is presented in \cite{Collins:2007ph}.
Here, we will use the unintegrated density obtained from the solution to the set of unified BFKL and DGLAP equations, which includes the small $x$ resummation effects. The full formalism, called the KMS approach,  was constructed  in \cite{Kwiecinski:1997ee} and here we only  review main elements of this approach.

\subsection{Equation for the unintegrated gluon density}

In the KMS approach \cite{Kwiecinski:1997ee}
 one constructs the evolution equation for the unintegrated gluon distribution function which includes the
leading order BFKL kernel with the kinematical constraint and the DGLAP part of the splitting function. 
The solution to the leading order BFKL equation is well known \cite{Lipatov:1985uk},  giving  very fast growth of the gluon
density with the decreasing value of $x$. It  can be recast into the symbolic form: 
\begin{equation}
\label{eq:bfklsol}
f(x,k^2) \sim x^{-\lambda} \; ,
\end{equation}
where  $\lambda\simeq  0.5$ is the so-called hard Pomeron intercept.
This fast growth was shown to be incompatible with 
the experimental data which exhibit an effective intercept $\lambda \simeq 0.3$.

The  next-to-leading corrections to the BFKL equation turned out to be very large \cite{Fadin:1998py,Ciafaloni:1998gs,Camici:1997ij}
and it became immediately  apparent that additional resummations are necessary. In the KMS approach one uses the
leading logarithmic approximation for the BFKL kernel but with substantial modifications. One of them is 
the kinematical constraint \cite{Andersson:1995jt,Kwiecinski:1996td} which accounts for a large portion
of the next-to-leading order corrections. It is important to stress that this constraint goes beyond the next-to-leading 
order in logarithms of $x$, and is responsible for  partial resummation of the small $x$ series \cite{Salam:1998tj}. Another modification in the KMS approach is a non-singular part in  $1/z$  of the splitting function $P_{gg}$ in the BFKL kernel in addition to the already included 
singular part. In a  series
of papers \cite{Ciafaloni:2002xf,Ciafaloni:2002xk,Ciafaloni:2007gf} (see also \cite{Altarelli:1999vw,Altarelli:2000mh,Altarelli:2001ji,Altarelli:2005ni,Vera:2005jt,White:2006yh})  it was shown in detail how these modifications generate  higher order terms  in the small $x$ expansion.

The final equation in the KMS approach, which takes into account all the modifications mentioned above, has the following form
\begin{widetext}
\begin{eqnarray}
\nonumber
 f(x, k^2) &=& \tilde{f}^{(0)} (x, k^2) \,+\,\overline{\alpha}_S (k^2)\, k^2\! \int_x^1 \frac{dz}{z} 
\int_{k_0^2} \frac{dk^{\prime 2}}{k^{\prime 2}} \left\{
\frac{f\left( {\displaystyle \frac{x}{z}}, 
k^{\prime 2} \right) \Theta \left({\displaystyle \frac{k^2}{z}} -
k^{\prime 2}\right) - 
f \left({\displaystyle \frac{x}{z}}, k^2\right)}
{| k^{\prime 2} - k^2 |} \; + \; \frac{f \left({\displaystyle
\frac{x}{z}}, k^2 \right)}{[4k^{\prime 4} 
+ k^4]^{\frac{1}{2}}} \right\} 
\\
 &+& \overline{\alpha}_S (k^2) \int_x^1 \frac{dz}{z}
\left(\frac{z}{6}
P_{gg} (z) - 1 \right ) \int_{k_0^2}^{k^2}  \frac{dk^{\prime 2}}
{k^{\prime 2}} f \left(\frac{x}{z}, k^{\prime 2} \right)
\,+\,\frac{\alpha_S (k^2)}{2\pi} \int_x^1 dz P_{gq} (z) \Sigma \left(\frac{x}{z}, k^2 \right) \; .
\label{eq:gluon}
\end{eqnarray}
\end{widetext}
where the strong coupling constant  $\bar{\alpha}_s\equiv \alpha_s N_c/\pi$. The first term on the r.h.s is the non-perturbative input to be specified below, while  the second term
contains the leading logarithmic BFKL kernel  with the kinematical constraint  given by the theta function. 
The third term contains the DGLAP splitting function $P_{gg}(z)$ without the singular term in $z$, and the last term is the contribution from the quark to gluon transition
with $\Sigma$ being the singlet quark distribution. The input function for this integral equation is chosen to be 
\begin{equation}
\label{eq:inputgluon}
 \tilde{f}^{(0)} (x, k^2) \; = \; \frac{\alpha_S
(k^2)}{2\pi} \int_x^1 dz\, P_{gg} (z)\, \frac{x}{z} g
\left(\frac{x}{z}, k_0^2 \right) .
\end{equation}
Note that the special form of the input is dictated by the fact that eq.~(\ref{eq:gluon}) only  involves $f (x, k^2)$ in
the perturbative 
domain, $k^2>k_0^2$, where $k_0^2$ is a non-perturbative cutoff taken to be equal 
$k_0^2 = 1 \; {\rm GeV}^2$.  
The gluon input (\ref{eq:inputgluon}) is provided by the conventional
gluon distribution $xg(x,k_0^2)$.  This guarantees  consistency with the DGLAP evolution equations since the input in both approaches is exactly of the same. In this way,
the necessity of parametrizing  the unintegrated gluon distribution in the non-perturbative regime, $k^2<k_0^2$, is avoided.
\subsection{Equation for the singlet quark  density}

\begin{figure*}
\centerline{
\includegraphics[width=0.75\textwidth]{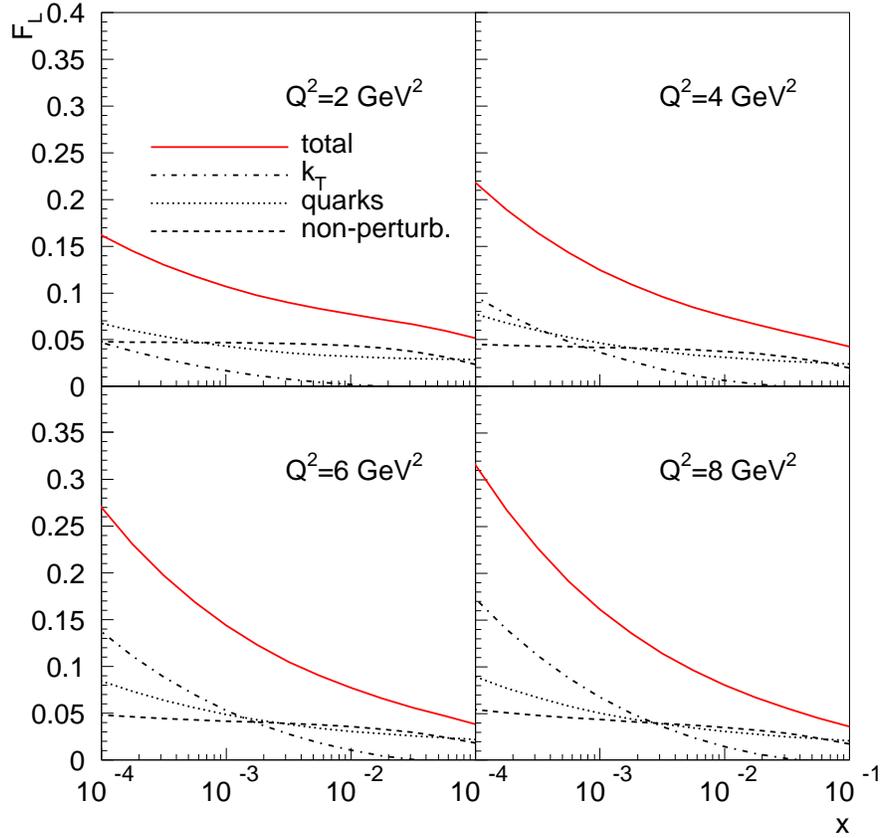}
}
\caption{$F_L$  from the  $k_T$ factorization approach. Dashed-dotted (black) line: contribution from the boson-gluon fusion box with gluon transverse momenta $k>k_0$; dashed (black) line: non-perturbative input from the gluons via collinear formula; dotted (black) line: contribution from the quarks. Solid (red) line is the sum of all contributions. Gluon kinematics is exact.}
\label{fig:ktsplitexact}
\end{figure*}

In the KMS approach the equation for the unintegrated gluon density was supplemented by the second equation for the quark density. 
These two equations formed the coupled system of equations (similarly to the DGLAP equations) 
for the functions $f(x,k^2)$ and $\Sigma(x,k^2)$.
The sea quark contribution was evaluated from the $k_T$ factorization theorem at small $x$,
\be
S_q^{(k)} (x, Q^2) \; = \; \int_x^1 \: \frac{dz}{z} \: \int
\: \frac{dk^2}{k^2} \: S^{\rm (box)}_q \: (z, k^2, Q^2) \: f \left
(
x_g, k^2 \right)
\label{eq:a13}
\ee
where $S_q^{\rm (box)}$ describes the quark box contributions, shown in Fig.~\ref{fig:qbox},  which implicitly includes the integration over the quark transverse momentum $\mbox{\boldmath $\kappa$}$.
 The explicit expression for (\ref{eq:a13}) reads
\begin{widetext}
\bea
\label{eq:a18}
S_q^{(k)} (x, Q^2) & = & \frac{Q^2}{4\pi^2} \: \int \:
\frac{dk^2}{k^4} \: \int_0^1 \: d\beta \: \int \: d^2
\kappa^\prime \alpha_S \left \{ [\beta^2 + (1 - \beta)^2 ] \:
\left (\frac{\mbox{\boldmath $\kappa$}}{D_{1 q}} \: - \:
\frac{\mbox{\boldmath $\kappa$} - \mbox{\boldmath $k$}}{D_{2q}}
\right )^2 \right . 
\nonumber \\
& + & [m_q^2 \: + \: 4Q^2 \beta^2 (1 - \beta)^2 ] \:
\left (\frac{1}{D_{1 q}} \: - \: \left . \frac{1}{D_{2q}} \right
)^2 \right \} \: f \left (\frac{x}{z}, k^2
\right ) \Theta \left(1 - \frac{x}{z} \right) 
\eea
\end{widetext}
where  the quantities $\kappa'$, $D_{1q},D_{2q}$ and $x_g$  were defined in Sec.~\ref{sec:flkt}.

The singlet quark momentum distribution  contains both the sea and  valence quarks.
The contribution to the singlet quark distribution was calculated differently depending on the region
of the transverse momenta. There are three regions of interest for $k^2$ and $\kappa^{\prime 2}$:
\begin{enumerate}
\item non-perturbative: $k^2,\kappa^{\prime 2} < k_0^2$
\item strongly ordered with low gluon transverse momenta: $k^2<k_0^2<\kappa^{\prime 2}$
\item perturbative:  $k^2 > k_0^2$\,.
\end{enumerate}

In the non-perturbative region, the sea contribution
is assumed to be dominated by the soft-Pomeron exchange.  This part is 
parametrized phenomenologically in the following  form 
\be
S^{(\rm soft)}(x) = S_u^{P}+S_d^{P}+S_s^P 
\label{eq:softpom}
\ee
with the soft pomeron contribution
\be
S_u^P=S_d^P=2 S_s^P = C_P x^{-0.08}(1-x)^8 \, .
\ee

The second contribution
comes from the region of  small transverse momenta of the gluon, $k^2<k_0^2<\kappa^{\prime 2}$.  In this region the strongly ordered approximation for the quark-gluon transition is applied and the relevant  contribution is given by the following formula
\be
S^{(\rm coll)}(x,Q^2) = \int_x^1 \frac{dz}{z}\, S_q^{\rm (box)}(z,k^2=0,Q^2) \,\frac{x}{z}g\!\left(\frac{x}{z},k_0^2\right)
\ee
where the on-shell approximation,  $k^2=0$, is applied to evaluate  $S_q^{\rm (box)}$. 
 
In the perturbative domain, $k^2>k_0^2$, the quark contribution is evaluated from the $k_T$ factorization formula.
The final expression for the singlet quark distribution is taken to be the sum of the contributions from the three discussed regions
\be
\Sigma \; = \; (S_{uds}^{(\rm soft)} \: + \: S_{uds}^{(\rm coll)} \: + \: S_{uds}^{(k)})
\: + \: (S_{c}^{(\rm coll)} + S_{c}^{(k)}) \: + \: V \; .
\label{eq:a19}
\ee
Note that for the charm evaluation  we did not use the soft contribution since we assume that charm is generated dynamically from gluons and that there is no soft or non-perturbative charm contribution.

Using the $k_T$ factorization and all the terms discussed above, one finds the
the final equation for the singlet distribution $\Sigma$ in the KMS approach: 
\begin{widetext}
\bea
\label{eq:z13}
\Sigma (x, k^2) & = & S^{(\rm soft)} (x) \: + \: \sum_q \int_{x}^a
\frac{dz}{z} \:
S_q^{\rm (box)} (z, k^{\prime 2} = 0, k^2; m_q^2) \frac{x}{z} \: g
\left(\frac{x}{z},
k_0^2 \right)
\: + \: V (x, k^2) 
\\\nonumber
&  + & \sum_q \int_{k_0^2}^\infty \frac{dk^{\prime 2}}{k^{\prime 2}} 
\int_x^1 \frac{dz}{z} \: S_q^{\rm (box)} (z, k^{\prime 2},
k^2; m_q^2) f \left (\frac{x}{z}, k^{\prime 2} \right)
\,+ \, \int_{k_0^2}^{k^2} \frac{dk^{\prime 2}}{k^{\prime 2}} \:
\frac{\alpha_S
(k^{\prime 2})}{2\pi} \int_x^1 dz \: P_{qq} (z)
S_{uds} \left(\frac{x}{z}, k^{\prime 2} \right )
\eea
\end{widetext}
where $S^{(\rm soft)}(x)$ is given by eq.~(\ref{eq:softpom}) and
the $uds$ subscript indicates that the additional $S
\rightarrow S$ term is only included for the light quarks.  Eqs.~(\ref{eq:z13}) and (\ref{eq:gluon})
form a set of equations to be solved  in the KMS approach.

The final formula  used for the calculation of the longitudinal structure function $F_L$ from the $k_T$ factorization formalism with the KMS approach reads:
\begin{widetext}
\begin{eqnarray} 
\nonumber
F_L(x,Q^2)&=& {Q^4\over \pi^2} \sum _qe_q^2 \int {dk^2\over k^4}\, \theta(k^2>k_0^2)\int_0^1 d\beta
 \int d^2{\tdm{\kappa^{\prime}}} \,
\alpha_s(\mu^2) \, \beta^2 (1-\beta)^2 \left({1\over D_{1q}} - {1\over D_{2q}}\right)^2  f(x_g,k^2) 
\\\nonumber
\\
&+&\frac{\alpha_s(Q^2)}{\pi}\left\{\frac{4}{3}\int_x^1\frac{dy}{y}
\left(\frac{x}{y} \right)^2F_2(y,Q^2) \right.
 +\left.\sum_qe_q^2\int_x^1\frac{dy}{y} \left(\frac{x}{y}
\right)^2
\left(1-\frac{x}{y} \right)yg(y,k_0^2)\right\}.
\label{flint2}
\end{eqnarray} 
\end{widetext}
where the cutoff for the gluon momentum $k_0^2= 1 \; {\rm GeV^2}$ and the  non-perturbative input was taken to be
\be
yg(y,k_0^2) = N (1-y)^{\beta}
\ee
with the parameters $N=1.57$ and $\beta=2.5$ which were found from a fit to the HERA data on the structure function $F_2$, using the same approach.
In the forthcoming, we will discuss in detail the relation of the $k_T$ factorization formula (\ref{flint2})  to those from  other approaches.

It is interesting to see what is the magnitude of the separate contributions to $F_L$ in eq.~(\ref{flint2}). In Fig.~\ref{fig:ktsplitexact} we show the breakdown of $F_L$ into the contributions from  gluons from the boson-gluon box (first term)), quarks (second term) and the non-perturbative gluon input (third term). 
The non-perturbative input stays nearly constant as a function of $x$ and $Q^2$. At low $x$ the dominant contribution is from the gluon density in the $k_T$ factorization framework. However, this contribution is small at $x>0.01$ and at small values of $Q^2$. This is due to the kinematic effects since  phase space for the gluon emissions shrinks in this regime. On the other hand, the quark contribution is non-negligible in the same regime.


\section{Relation of the $k_T$  factorization to other approaches}
\label{sec:colldipole}

\subsection{Relation to the collinear factorization approach}

The standard collinear factorization formula for the longitudinal structure function
reads
\begin{widetext}
\begin{equation}
F_L(x,Q^2) \; = \;  \frac{\alpha_s(Q^2)}{\pi} \bigg[ 2\sum_q e_q^2 \int_x^1 \frac{dy}{y} \, \left(\frac{x}{y}\right)^2 \, \left( 1-\frac{x}{y}\right)\, yg(y,Q^2) + 
\frac{4}{3} \int_x^1 \frac{dy}{y} \left(\frac{x}{y}\right)^2 F_2(x,Q^2)\bigg] \; ,
\label{eq:flsimpleonshell}
\end{equation}
\end{widetext}
where quark masses in this formula are neglected.
Thus, $F_L$ has two contributions: originating from quarks and proportional to $F_2$, and from gluons and proportional to  the  integrated gluon distribution $g(y,Q^2)$.


\begin{figure*}[t]
\centerline{
\includegraphics[width=0.44\textwidth]{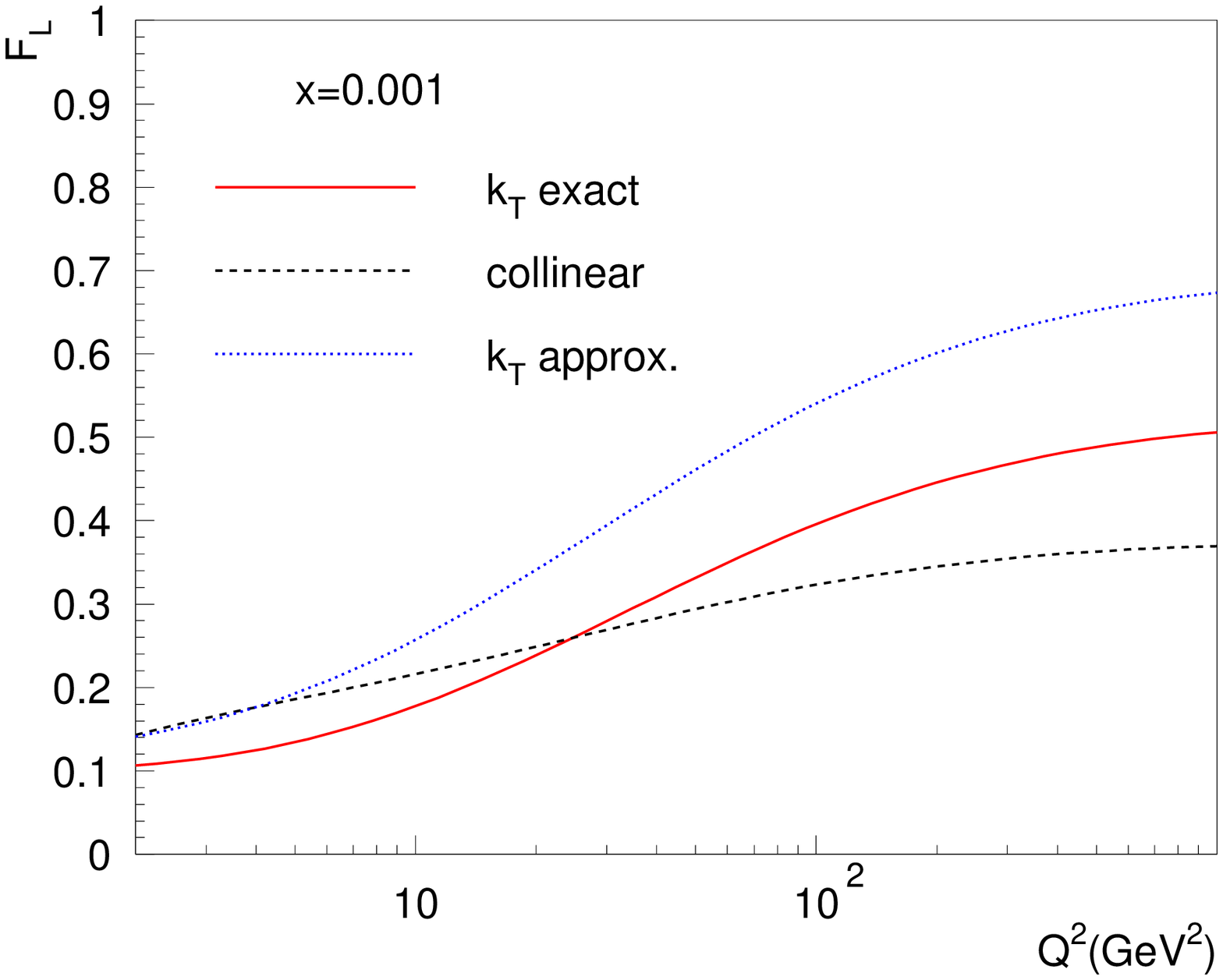}
\includegraphics[width=0.44\textwidth]{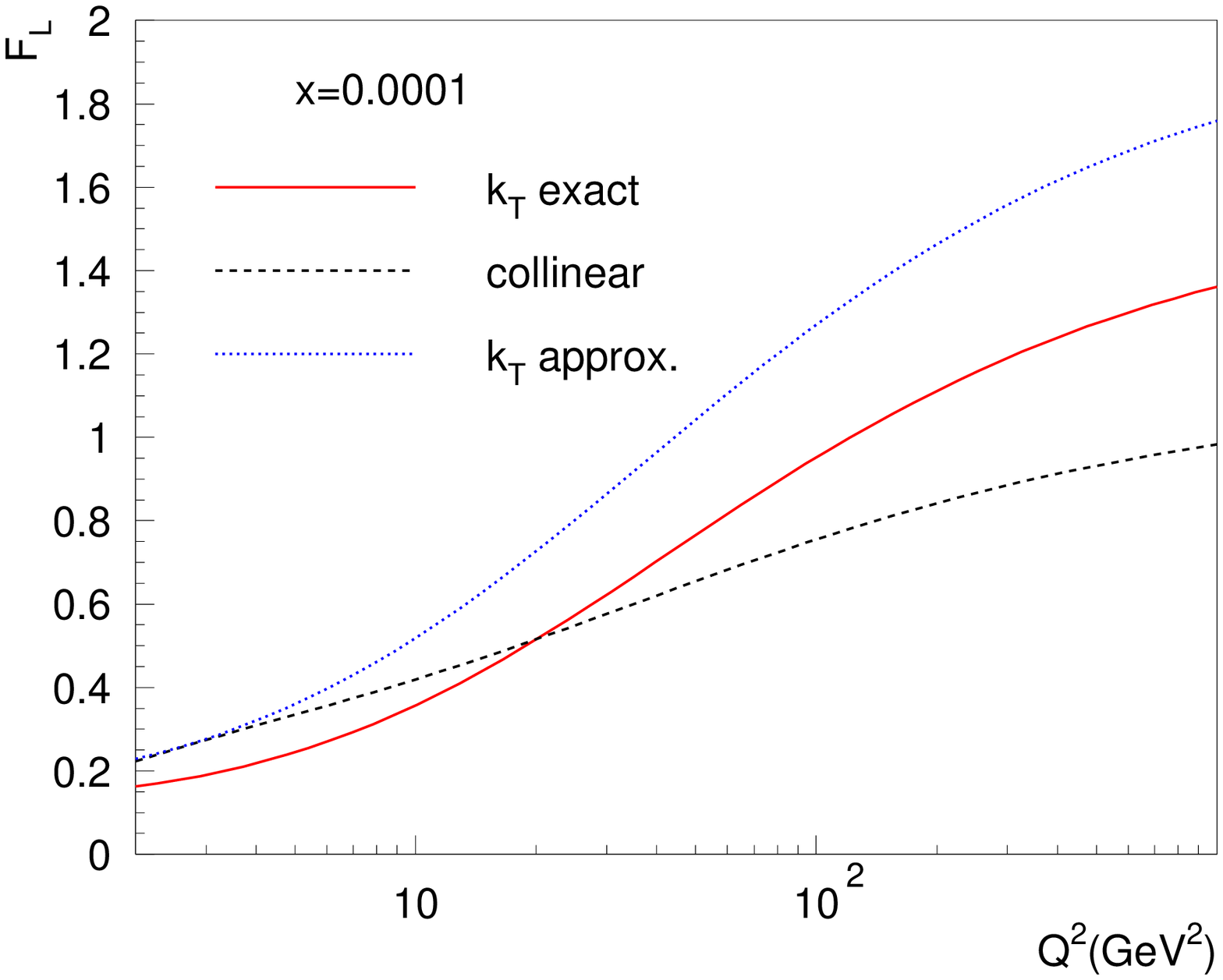}
}
\caption{Comparison between the $k_T$ factorization and collinear factorization  predictions for $F_L$ as a function of $Q^2$ for two values of $x$: $10^{-3}$, (left) and $10^{-4}$ (right). The solid lines are  the $k_T$ factorization prediction with exact kinematics, eq.~(\ref{flint2}),  while  the dotted lines correspond to the dipole approximation, that is $x_g\to x$ in the gluon density. The dashed lines show the   
collinear factorization predictions, eqs.~(\ref{eq:flsimpleonshellm2},\ref{eq:flsimpleonshellm3}).}
\label{fig:flfunq2}
\end{figure*}
It is instructive to illustrate that the  on-shell limit of the $k_T$ factorization formula (\ref{flint}) is compatible with formula (\ref{eq:flsimpleonshell}). By the on-shell limit we mean the approximation in which the transverse momentum of the gluon $k^2$ is much smaller than the virtuality of the photon, $k^2\ll Q^2$.
To this aim, we start by expanding the expression under the integral  in eq.~(\ref{flint}),
\begin{equation}
\label{eq:flintegrand}
\frac{1}{2}\left({1 \over D_{1q}} - {1 \over D_{2q}}\right)^2,
\end{equation}
in powers of  $k^2/Q^2$. We retain only the leading term, proportional to $k^2$, and drop all the higher powers of   $k^2$.
Explicit expressions for the denominators $D_{1q}$ and $D_{2q}$ read
\begin{eqnarray}\nonumber
D_{1q} &=&{\kappa'}^2+2(1-\beta)\tdm{\kappa'} \cdot \tdm{k}
+(1-\beta)^2 k^2
\\\nonumber
&+&\beta(1-\beta)Q^2+m^2_q
\\\nonumber
D_{2q} &=&\kappa '^2-2\beta \tdm{\kappa'} \cdot \tdm{k}
+\beta ^2 k^2+\beta(1-\beta)Q^2+m^2_q\,.\nonumber
\label{d1d2}
\end{eqnarray}
After expanding the denominators in $k^2$ we obtain 
\begin{eqnarray}
{1 \over D_{1q}} &=&{1 \over D_q}-{2(1-\beta)\tdm{\kappa'} \cdot \tdm{k} \over D_q^2}
+O(k^2)\;,\nonumber \\
{1 \over D_{2q}} &=&{1 \over D_q}+{2\beta\tdm{\kappa'} \cdot \tdm{k} \over D_q^2}
+O(k^2)\;,\nonumber
\label{rozw}
\end{eqnarray}
where
\begin{equation}
D_q=\kappa'^2+\beta(1-\beta)Q^2+m^2_q
\end{equation}
is independent of the gluon transverse momentum ${\tdm{k}}$. Here, we need to keep only terms linear in $k$.
Therefore, expression (\ref{eq:flintegrand}) to the first order in $k^2$ reads
\begin{equation}
\left({1 \over D_{1q}} - {1 \over D_{2q}}\right)^2=
{4\cos^2\!\phi \, \kappa^{\prime 2} \, k^2 \over D_q^4}\;. 
\end{equation}
The integration 
 $d^2\kappa'$ can be written as  ${1 \over 2}d\kappa'^2 d\phi$
and one can perform the azimuthal integration over the angle $\phi$.
The dependence on $k^2$ is now only in the unintegrated gluon distribution. Since we have 
assumed the strong ordering in the transverse momenta, we can easily perform this integration using the following definition
\begin{equation}
yg(y,\mu^2) \equiv \int^{\mu^2}{dk^2 \over k^2}f(y,k^2)
\label{intglu}
\end{equation}
where the scale $\mu^2\sim Q^2$. Note that, formally the integration over the $k$ in eq.~ (\ref{flint}) is over all scales. However,
we expand the $k_T$ factorization formula for small values of $k^2/Q^2$ and therefore, assuming that transverse momenta are small.
Using relation (\ref{intglu}), we can rewrite the approximate form as
\beeq\nonumber
F_L^{\rm (on-shell)}(x,Q^2) & = &  2{Q^4 \over \pi}\sum_q e^2_q\int_0^1d\beta\int d\kappa^{\prime 2} \, 
\alpha_s(Q^2)
\\
&\times& \beta^2(1-\beta)^2\,\frac{\kappa^{\prime 2}}{D_q^4}\, yg(y,\mu^2)
\label{offshell1}
\eeeq
where now
\begin{equation}
\label{eq:yargument}
y\equiv x\left(1+{\kappa'^2+m^2_q \over \beta(1-\beta)Q^2}\right)\; ,
\end{equation}
since  we have dropped the ratio $k^2/Q^2$ in $x_g$, see eq.~(\ref{eq:xgluon}).
It is convenient to change the integration variables in eq.~(\ref{offshell1}) from  $\kappa'^2$
to $y$, 
\begin{eqnarray}
\label{kappa}
\kappa^{\prime 2}\eq\beta(1-\beta)Q^2\left({y \over x}-1\right)-m_q^2 \; ,
\\
D_q&=&\beta(1-\beta)Q^2{y \over x}\;,
\end{eqnarray}
and carefully set the integration limits.
Using relation  (\ref{kappa}) we can write the inequality
\begin{equation}
\beta(1-\beta)Q^2\left({y \over x}-1\right)-m_q^2>0\;,
\end{equation}
and since $1>\beta>0$, we have
\begin{equation}
{1 \over 4}>\beta(1-\beta)>\frac{m_q^2 x}{Q^2(y-x)}\;.
\label{lim}
\end{equation}
>From inequality (\ref{lim}) we obtain the lower  limit for $y$
\begin{equation}
y>x\left(1+{4m^2_q \over Q^2}\right)\;.
\end{equation}
It is now convenient to make another change of variables: 
\be
\beta=\frac{1}{2}+\lambda\,,
\ee
and using inequality (\ref{lim}), we  finally obtain
\begin{equation}
-\sqrt{\frac{1}{4} - \frac{m_q^2 x}{Q^2(y-x)}}<\lambda<
\sqrt{\frac{1}{4} - \frac{m_q^2 x}{Q^2(y-x)}}\;.
\end{equation}
As a result, we obtain the following expression for the on-shell limit of the $k$ factorization formula
\begin{equation}
F_L^{\rm (on-shell)}(x,Q^2) = 2 \sum_q e_q^2\left[J_q^{(1)}-2\,\frac{m_q^2}{Q^2}J_q^{(2)}\right]
\label{eq:flsimpleonshellm1}
\end{equation}
where
\bea\nonumber
J_q^{(1)}&=&\frac{\alpha_s}{\pi} \int_{\bar{x}_q}^1 \frac{dy}{y} \bigg(\frac{x}{y}\bigg)^2 \bigg(1-\frac{x}{y}\bigg)
\\
&\times& \sqrt{1-\frac{4m_q^2 x}{Q^2 (y-x)}} \, yg(y,Q^2) \; ,
\label{eq:flsimpleonshellm2}
\eea
and 
\bea\nonumber
J_q^{(2)}&=&\frac{\alpha_s}{\pi} \int_{\bar{x}_q}^1 \frac{dy}{y} \bigg(\frac{x}{y}\bigg)^3 
\\
&\times&
\ln \left[\frac{1+\sqrt{1-\frac{4m_q^2 x}{Q^2 (y-x)}} }{1-\sqrt{1-\frac{4m_q^2 x}{Q^2 (y-x)}} }\right]\, yg(y,Q^2)
\label{eq:flsimpleonshellm3}
\eea
with the lower cutoff on the integration equal to 
\be
\bar{x}_q \, = \, x\bigg(1+\frac{4m_q^2}{Q^2}\bigg) \; .
\ee

Formula (\ref{eq:flsimpleonshellm1}) together with eqs.~(\ref{eq:flsimpleonshellm2}) and (\ref{eq:flsimpleonshellm3})
is the on-shell approximation derived from the $k_T$ factorization in the presence of  quark masses.
In this derivation we also assumed that the argument of the coupling constant is equal to the external scale $\mu^2\simeq Q^2$.
It is straightforward to verify that the above expressions coincide with the gluonic contribution of the  standard massless collinear formula (\ref{eq:flsimpleonshell}) in the case when the quark masses
vanish. Therefore, the collinear formula arises as a leading twist part of the $k_T$ factorization formula, and the second term in eq.~(\ref{eq:flsimpleonshellm1}) contains a part of the higher twist contribution proportional to ${1}/{Q^2}$.

Let us emphasize that in order to obtain the limit of the $k_T$ factorization formula consistent  with the collinear factorization, it was crucial to take
the exact kinematics for the argument of the gluon density, $x_g$, see formulae (\ref{flint}), (\ref{eq:xgluon}) and (\ref{eq:yargument}). 
The numerical relevance of the above substitution in the $k_T$ factorization formula is shown in Fig.~\ref{fig:flfunq2}.  In this figure we show the $Q^2$ dependence of $F_L$ obtained within the $k_T$ factorization approach with and without the exact kinematics, as well as the computation within the collinear approach.
The differences  between the collinear and $k_T$ factorization  approaches are not significant although
there is distinctive difference in the slope of the $Q^2$ dependence. On the other hand, the differences due to the exact gluon kinematics are quite substantial and persist even in the small $x$ regime.

\begin{figure*}[t]
\centerline{
\includegraphics[width=0.44\textwidth]{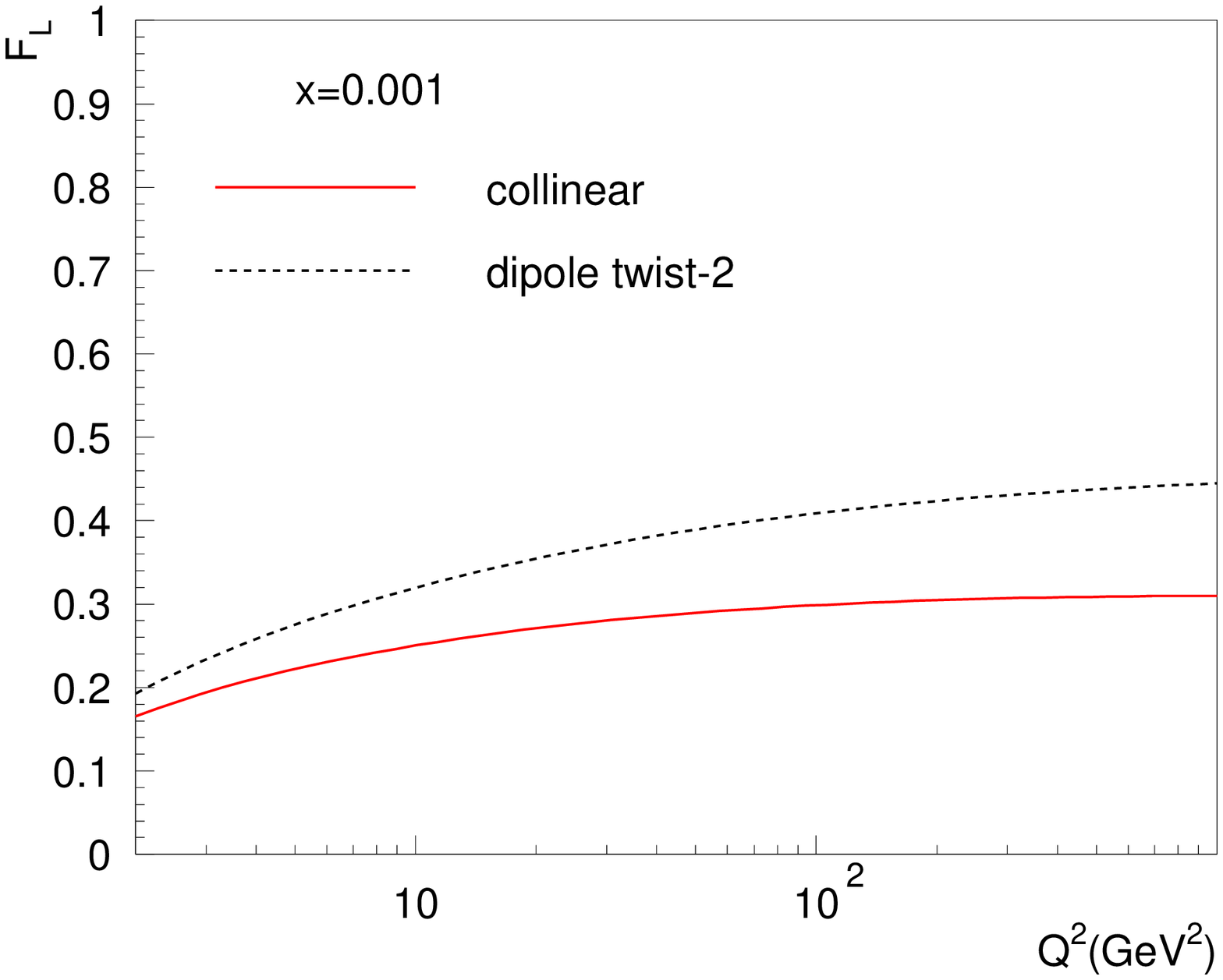}
\includegraphics[width=0.44\textwidth]{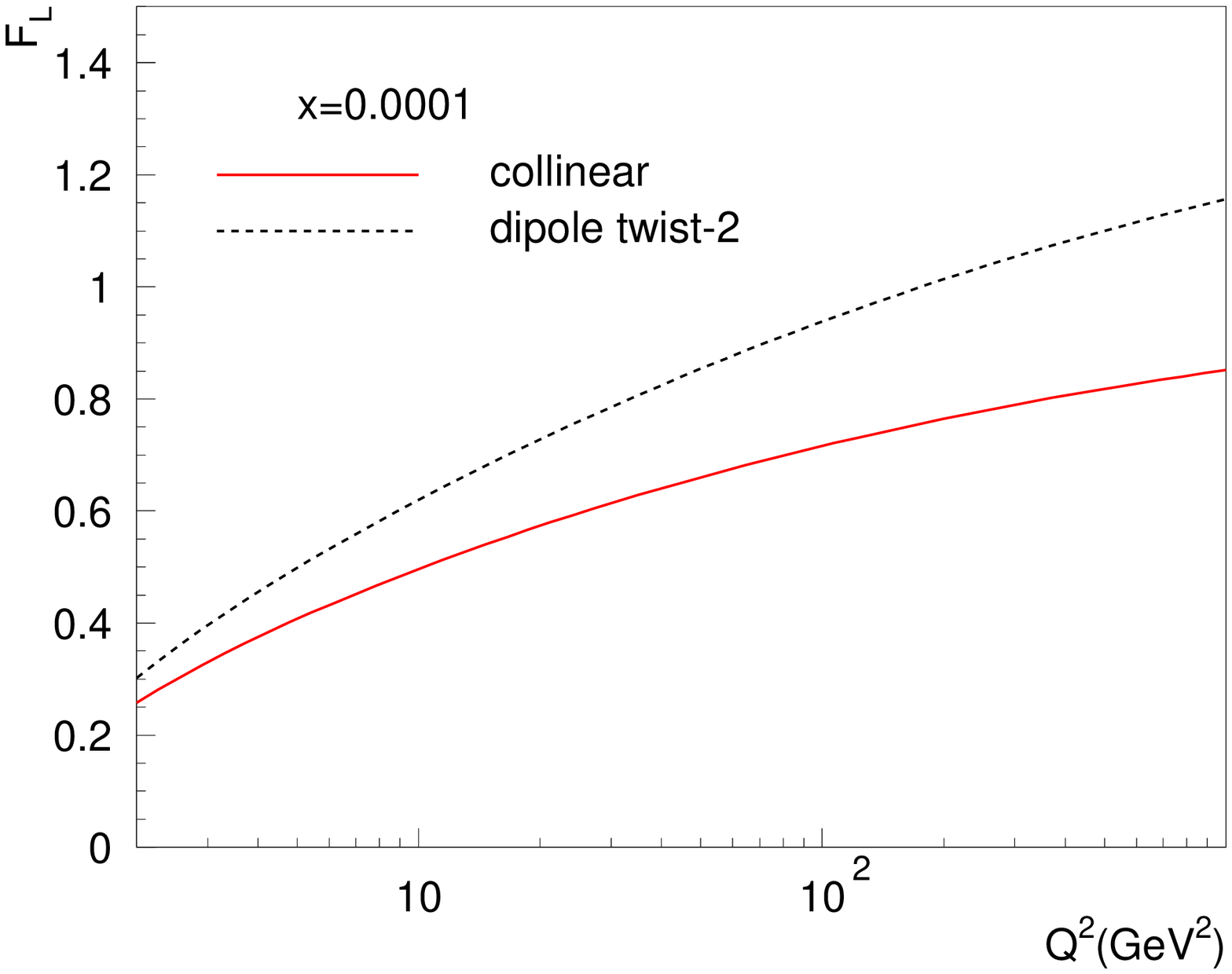}
}
\caption{Comparison between the collinear factorization formula (\ref{eq:flsimpleonshell}) with only the gluon term 
and massless quarks (solid lines)  and the dipole twist-2 predictions (dashed lines) as a function of $Q^2$ for two values of $x$: $10^{-3}$ (left) and  $x=10^{-4}$ (right). 
}
\label{fig:flfunq2twist2}
\end{figure*}

\subsection{Relation to the dipole approach}
\label{sec:colldipole:b}

The dipole representation for the inclusive cross section can be computed from the $k_T$ factorization formula. It is obtained after the Fourier transformation of expression (\ref{flint}) from  the space of quark transverse momenta $\tdm{\kappa}$   into the the space of the transverse coordinates, $\bf{r}$.  It is important to note that one also needs to perform the small $x$ approximation in the argument of the gluon density in formula (\ref{flint}),
\be\label{eq:xgtox}
x_g \rightarrow x\,.
\ee 
This is obviously justified only in the limit of very small $x$.
In this way the Fourier integrals over the $\tdm{\kappa}$ variable in (\ref{flint}) can be easily performed.

In the case of the longitudinal structure function, 
\be
F_L=\frac{Q^2}{4\pi^2\alpha_{em}}\,\sigma_L\,,
\ee
where the longitudinally polarized photon-proton cross section $\sigma_L$ reads
\beeq\nonumber
\label{eq:sigmaldipole}
\sigma_L\eq \frac{\alpha_{em}}{\pi} \sum_q e_q^2 \int d^2{\tdm{r}} \int_0^1 d\beta \, 4 Q^2 \beta^2\, (1-\beta)^2
K_0^2({\overline{Q}r})
\\
&\times& \int \frac{d^2 {\tdm{k}}}{k^4}\, \alpha_s f(x,k^2) (1-e^{-i {\tdm{r}} \cdot {\tdm{k}}})(1-e^{i {\tdm{r}} \cdot {\tdm{k}}}) \; .
\eeeq
Here $\tdm{k}$ is the gluon transverse momentum, the variable 
\be
\overline{Q}^2={\beta(1-\beta)Q^2+m_q^2}\,,
\ee 
 $K_0$ is the Bessel function and $\tdm{r}$ is the transverse size of the $q\bar{q}$ pair.  
The expression with the integral over $\tdm{k}$ in eq.~(\ref{eq:sigmaldipole}) is  proportional to the dipole-proton cross section,
$\hat\sigma$, which characterizes the interaction of the $q\bar{q}$ color dipole with the proton:
\be
\hat{\sigma}(x,{\tdm{r}}) \equiv \frac{2\pi}{3} \int \frac{d^2 {\tdm{k}}}{k^4} \alpha_s f(x,k^2) (1-e^{-i {\tdm{r}} \cdot {\tdm{k}}})(1-e^{i {\tdm{r}} \cdot {\tdm{k}}})\,.
\label{eq:dipolecrosssection}
\ee
Notice that such a Fourier transformation is only possible if the substitution $(\ref{eq:xgtox})$ is done. Otherwise,
by including the exact kinematics in the argument of the gluon distribution,  the transverse size of the quark-antiquark dipole is no longer conserved \cite{Bialas:2000xs,Bialas:2001ks,Bartels:2002cj}.

In \cite{Bartels:2000hv} a systematic analysis of the twist expansion (i.e. the expansion in powers of $1/Q^2$) in the dipole model approach was performed.  Using the GBW saturation model \cite{Golec-Biernat:1998js} for the dipole-proton cross section, a complete hierarchy of the twist series has been established.
The analysis has been extended in \cite{Motykaprep} to a saturation model which includes 
the DGLAP evolution \cite{Bartels:2002cj}. 
The higher twist terms are proportional to the nonlinear terms in  the gluon density. Consequently,
the leading twist part is the term which is linear in the gluon density.  
The leading twist-2 part in the dipole picture in the case of the cross section for the longitudinally polarized photon reads \cite{Bartels:2000hv}
\be
 \sigma_L^{\rm (twist-2)} =  \frac{4\pi \alpha_{em}}{3}\sum_q e_q^2\, \frac{\alpha_s xg(x,Q^2)}{Q^2}\; ,
\ee
or for the longitudinal structure function
\begin{equation}
F_L^{\rm (twist-2)} =  \frac{1}{3 \pi}\sum_q e_q^2 \, \,\alpha_s xg(x,Q^2) \, .
\label{eq:twist2}
\end{equation}
 
As we noted above, the dipole picture can be recovered from the $k_T$ factorization formula (\ref{flint})
by taking the high energy limit and making the substitution (\ref{eq:xgtox}).  In order to recover the leading twist-2 expression in the high energy limit, it is therefore enough  to consider  the gluon part of the collinear formula (\ref{eq:flsimpleonshell}) and replace the integrated gluon density by
\be
yg(y,Q^2) \rightarrow xg(x,Q^2)\,,
\ee
which allows to pull it outside the integral. The integral over $y$ can be then performed exactly 
\be\nonumber
\int_x^1 \frac{dy}{y} \left( \frac{x}{y}\right)^2\left(1-\frac{x}{y}\right) \; = \; \frac{1}{6}+\frac{1}{6} x^2 (2x-3)\simeq \frac{1}{6} \; ,
\ee
where we assume that $x\ll 1$. Combining the above result with eq.~(\ref{eq:flsimpleonshell}), we obtain the  twist-2 contribution to $F_L$ in the small $x$ approximation given by eq.~(\ref{eq:twist2}).
Therefore, the leading twist term from the dipole approach is actually an approximation to  the leading twist collinear factorization formula (\ref{eq:flsimpleonshell}). In Fig.~\ref{fig:flfunq2twist2} we demonstrate numerically that the values of $F_L$ obtained from eq.~(\ref{eq:twist2}) are larger than those from the collinear factorization formula.

Summarizing, we have demonstrated how the collinear and dipole approaches are related to the $k_T$ factorization formula (\ref{flint}). The collinear formula is obtained upon expanding the $k_T$ factorization formula in powers of $k^2/Q^2$ and retaining the lowest order in this expansion. The exact gluon kinematics has to be taken into account in that procedure. On the other hand, the dipole approach is obtained from the $k_T$ factorization expression in the limit when $x$ is very small, which amounts  to approximating the gluon longitudinal momentum fraction $x_g$ by the Bjorken $x$.


\section{Comparison with the HERA data}
\label{sec:numerics}

We start our numerical analysis by comparing the calculations of the longitudinal structure function $F_L$ performed within the collinear formalism and the $k_T$ factorization formalism with and without the exact kinematics. To be precise, for the collinear calculation we use the formulae (\ref{eq:flsimpleonshellm1},\ref{eq:flsimpleonshellm2},\ref{eq:flsimpleonshellm3}) with the light quark masses set to zero and the charm quark mass equal to $m_c=1.5 \; {\rm GeV}$. There is also quark contribution in the form of the second term in Eq.~(\ref{eq:flsimpleonshell}) which is proportional to $F_2$.
We emphasize that we have  fitted only the $F_2$ data, thus the calculations for $F_L$ are then the absolute predictions.

\begin{figure*}
\centerline{
\includegraphics[width=0.95\textwidth]{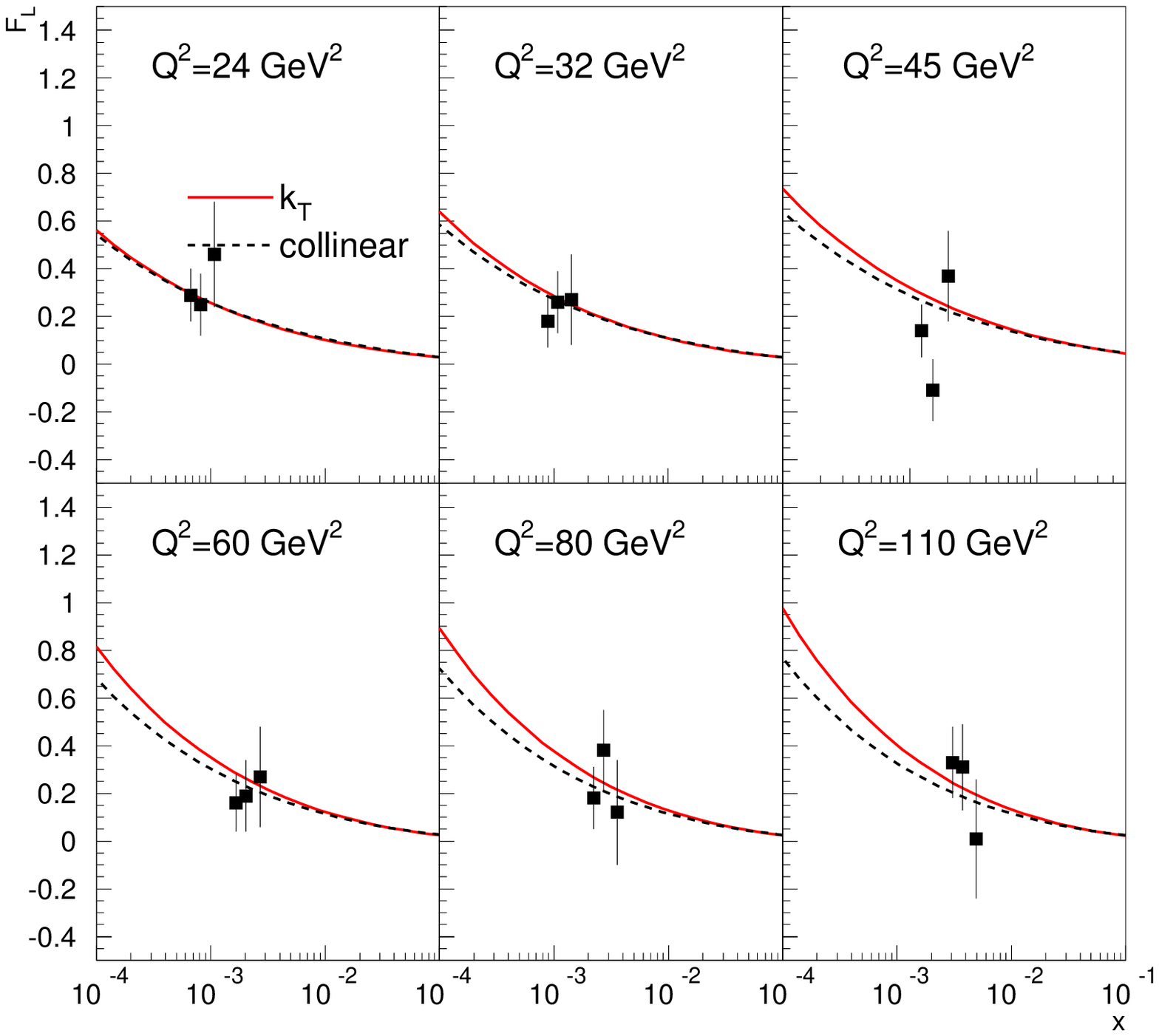}
}
\caption{Comparison between the collinear and the $k_T$ factorization calculations with exact gluon kinematics. 
The data are from the ZEUS experiment \cite{ZEUSFL}.
The light quarks $u,d,s$ are treated as massless, the charm quark mass is set to be $m_c=1.5 \, {\rm GeV}$. The solid (red) line denotes the calculation using the $k$ factorization with exact kinematics, the black (dashed) line is 
calculation using the collinear factorization with the massive charm quark. }
\label{fig:flktcollzeus}
\end{figure*}

\begin{figure*}
\centerline{
\hspace*{3cm}
\includegraphics[width=0.95\textwidth]{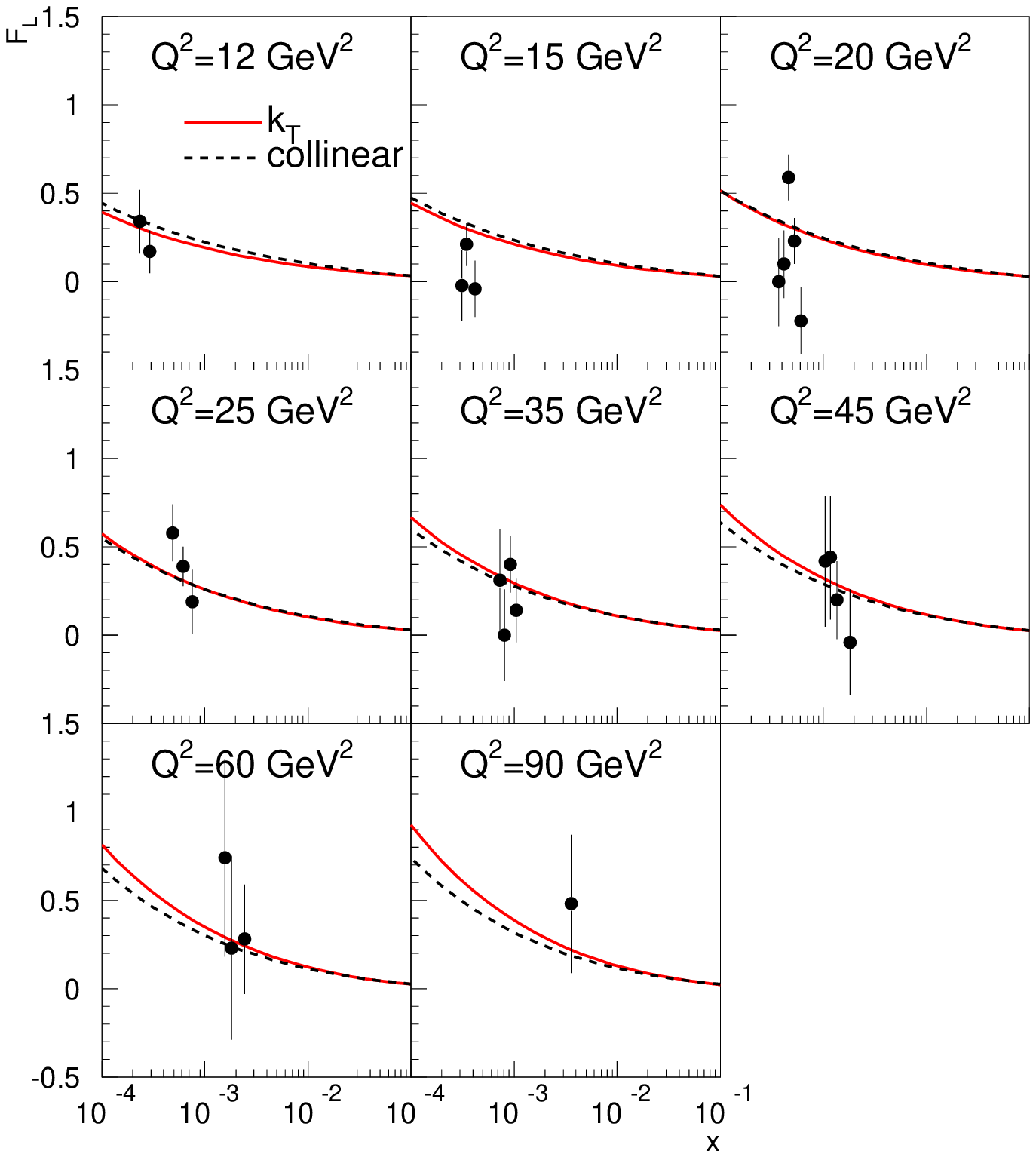}
}
\caption{Comparison between the collinear and the $k_T$ factorization calculations with exact gluon kinematics. The data are from H1 experiment \cite{:2008tx}. 
 The light quarks $u,d,s$ are treated as massless, the charm quark mass is set to be $m_c=1.5 \; {\rm GeV}$. The solid-red line denotes the calculation using the $k_T$ factorization with exact kinematics, the black-dashed line is calculation using the collinear factorization with the massive charm quark. }
\label{fig:flktcollh1}
\end{figure*}

In Figs.~\ref{fig:flktcollzeus} and \ref{fig:flktcollh1} we show   the calculations obtained using the collinear approach and the results from the $k_T$ factorization formalism with  the exact kinematics.  The $F_L$ structure function is plotted as a function of $x$ in bins of $Q^2$.
The data shown in Figs.~\ref{fig:flktcollzeus}  and Fig.~\ref{fig:flktcollh1} are from ZEUS \cite{ZEUSFL} and H1 \cite{:2008tx} experiments respectively. The agreement between the experimental data and our calculations  is good. In the case of the $k_T$ factorization with the exact kinematics the  results are rather close to the ones obtained from the collinear approach. For the kinematics range of the HERA data we do not see any significant differences  for $F_L$ between these two different factorization schemes. The only regions where the results differ is the region of  very small $x$ and high $Q^2$, where the $k_T$ factorization with exact kinematics tends to give higher values, and the region of small $Q^2$ (below $10 \; {\rm GeV^2}$) where the $k_T$ factorization-based approach falls below collinear one, compare Fig.~\ref{fig:flfunq2}.
This result is consistent with  the previous observations
\cite{Badelek:1996ap} which show that  the significant difference between the high energy and the collinear factorizations is more pronounced  for the transverse structure function \cite{Kwiecinski:1997ee}. 

\begin{figure*}
\centerline{\hspace*{3cm}
\includegraphics[width=0.95\textwidth]{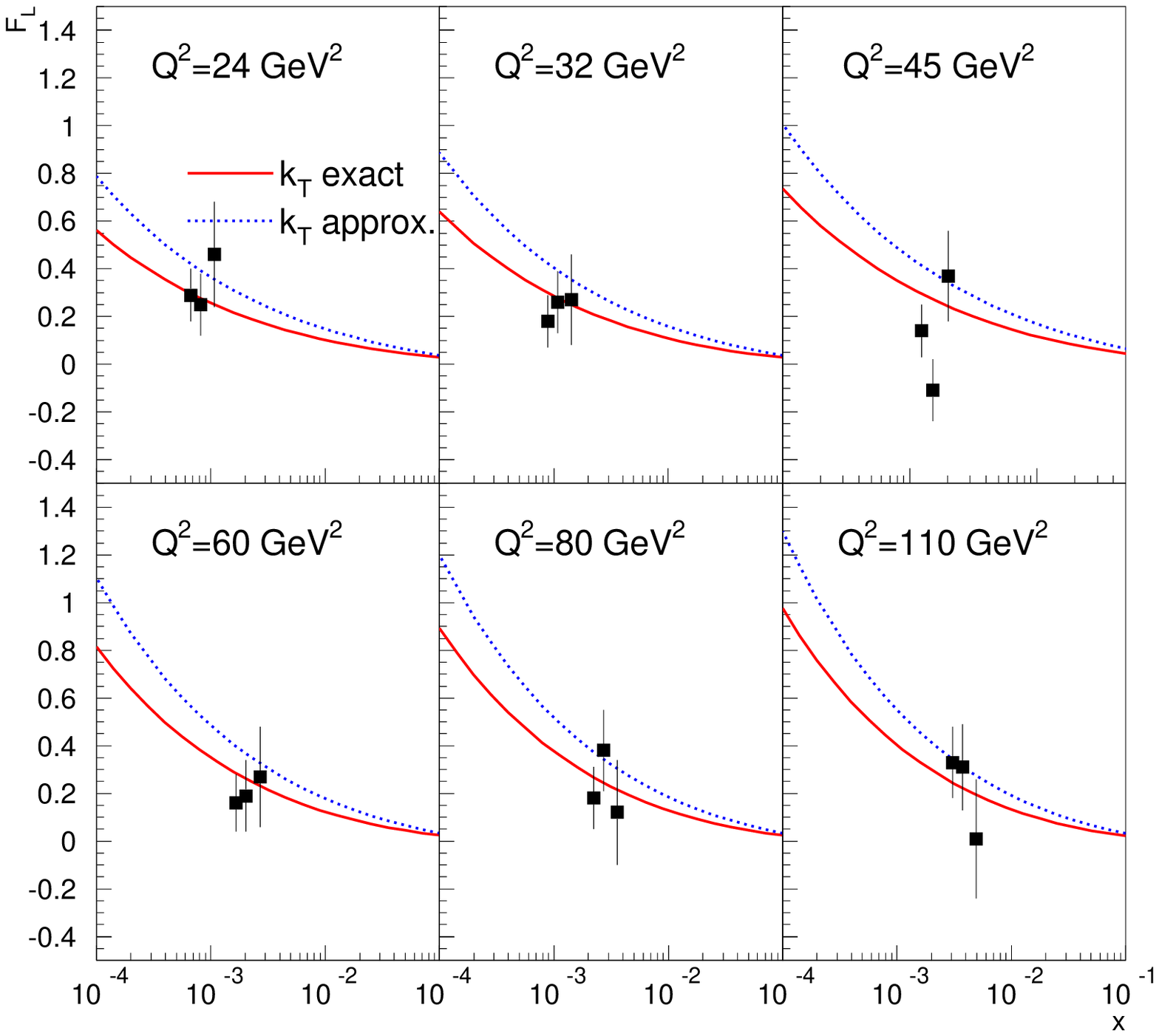}
}
\caption{Comparison between the  exact and the approximate (dipole like) kinematics in the $k_T$ factorization formula. 
The data are from the ZEUS experiment \cite{ZEUSFL}.
  The light quarks $u,d,s$ are treated as massless, the charm quark mass is set to be $m_c=1.5 \, {\rm GeV}$. The solid (red) line denotes the calculation using the $k_T$ factorization with exact kinematics; the dotted ( blue)  line is calculation using the $k_T$ factorization with the approximate kinematics.}
\label{fig:flktcollzeusea}
\end{figure*}

\begin{figure*}
\centerline{\hspace*{3cm}
\includegraphics[width=0.95\textwidth]{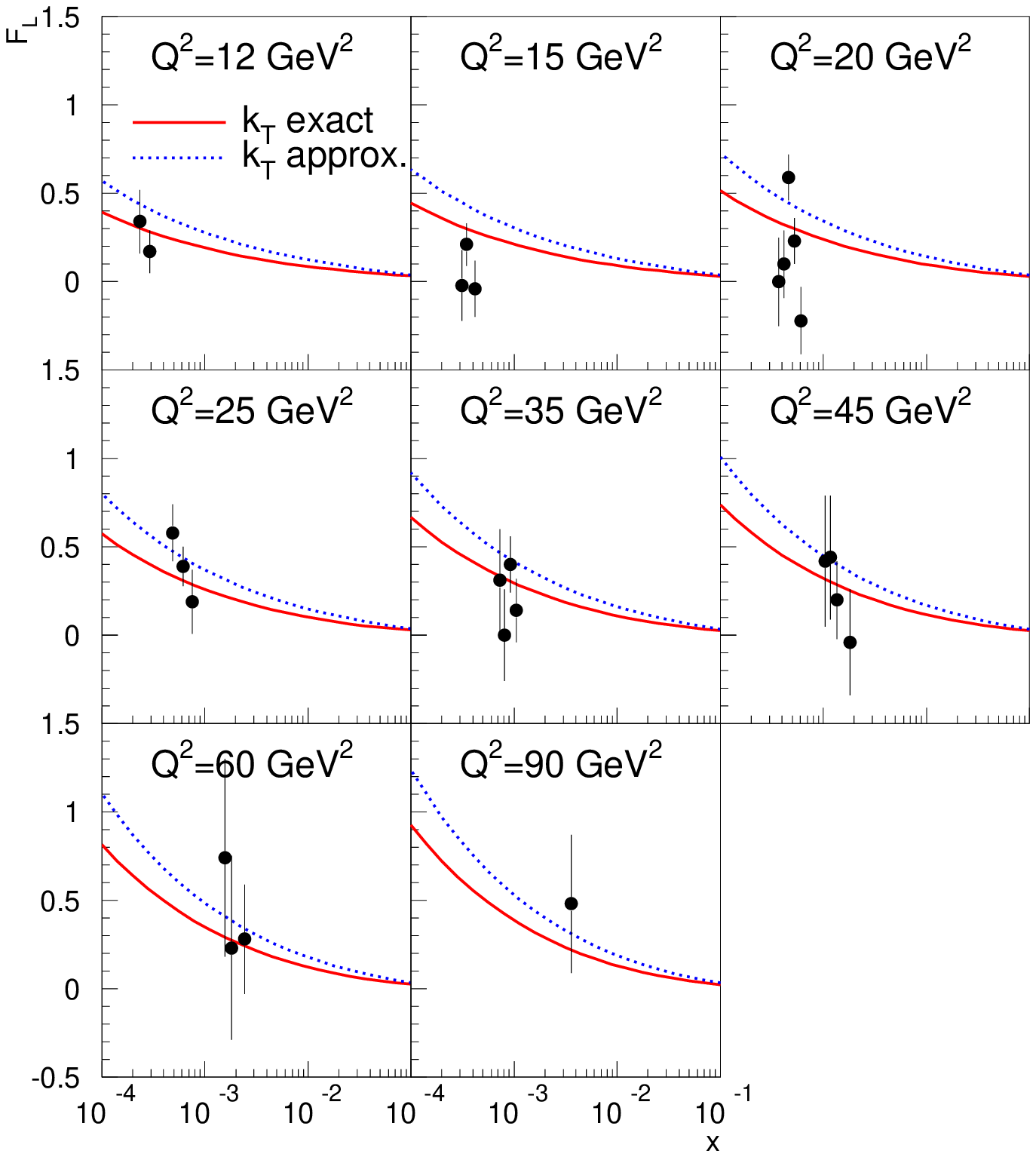}
}
\caption{Comparison between the  exact and the approximate (dipole like) kinematics in the $k_T$ factorization formula. 
 The data are from H1 experiment \cite{:2008tx}. 
 The light quarks $u,d,s$ are treated as massless, the charm quark mass is set to be $m_c=1.5 \; {\rm GeV}$. The solid, red line denotes the calculation using the $k_T$ factorization with exact kinematics; the blue, dotted line is calculation using the $k_T$ factorization with the approximate (dipole-like) kinematics.}
\label{fig:flktcollh1ea}
\end{figure*}

We stress though  the importance of the exact kinematics in the evaluation of the gluon density. The collinear and $k_T$ factorization approaches give very similar results only in the case when the gluon density is evaluated at $x_g$ in the $k_T$ factorization formula. In Figs.~\ref{fig:flktcollzeusea} and \ref{fig:flktcollh1ea} we show also the  calculation where in the $k_T$ factorization the argument of the gluon density equals the Bjorken $x$.
Clearly, the results which do not take into account the exact kinematics  are much higher  than those with the exact kinematics. This is understandable  as we are taking into account that finite energy has been used for the production of the $q\bar{q}$ pair, and   as a result the argument of the gluon density $x_g>x$ . We see that the differences are quite pronounced, they are typically  larger than the differences between the collinear and the $k_T$ factorization with the exact kinematics.
The differences are also visible in the plots of $F_L$ as a function of $Q^2$. It is interesting that the differences do not seem to vanish as a function of $Q^2$. This difference can be of course accounted for by changing the  gluon density.  The results for the  $F_L$ and $F_2$ structure functions can be made
consistent within the two calculations (with and without exact kinematics) at the expense of having different normalizations for the gluon density.

We have checked that the contribution from the $k_T$ factorization which is proportional to the gluon density is about $2-4$ times smaller with the exact kinematics than the approximate calculation.
 We have found that the approximate kinematics yields similar results when $x_g \simeq 5.7 \, x$
with the proportionality coefficient being the slowly varying function of $Q^2$.

\begin{figure*}
\centerline{\hspace*{3cm}
\includegraphics[width=0.95\textwidth]{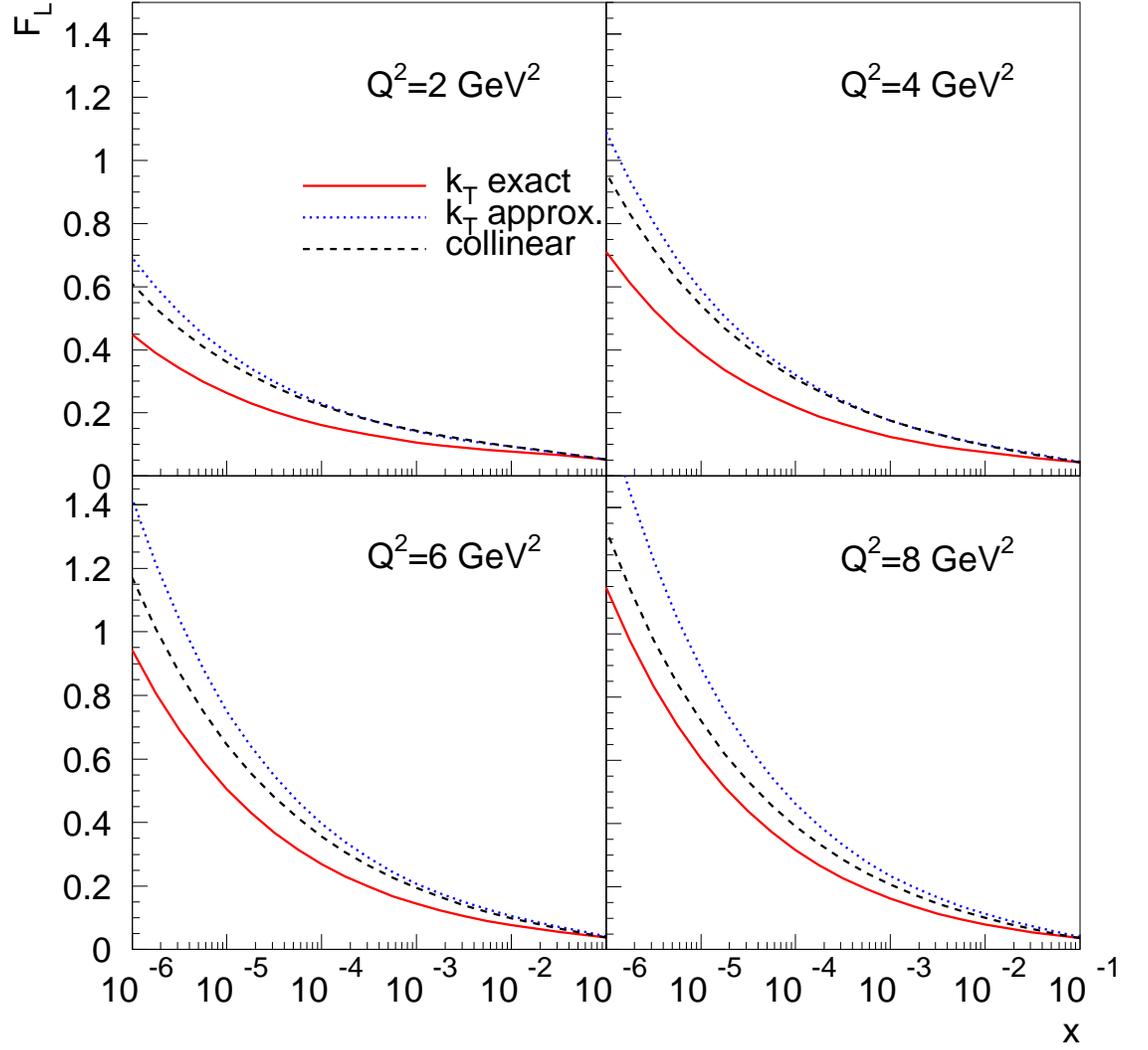}
}
\caption{ Comparison between the collinear and the $k_T$ factorization calculations for low values of $Q^2$.
The range in $x$ covers the LHeC kinematic range. 
  The light quarks $u,d,s$ are treated as massless, the charm quark mass is set to be $m_c=1.5 {\rm GeV}$. The solid (red) line denotes the calculation using the $k_T$ factorization with exact kinematics;  the  dotted (blue) line is calculation using the $k_T$ factorization with approximate kinematics and the dashed (black) line is the calculation using the  collinear factorization. }
\label{fig:ktcolllowq2}
\end{figure*}

In Fig.~\ref{fig:ktcolllowq2} we present the comparison of the $k_T$ and collinear factorizations for
low values of $Q^2\approx 2-8\;{\rm GeV^2}$.   We see that the two computations differ more in this region. The lowest values are given by the $k_T$ factorization approach with the  exact kinematics while the highest values are given by the calculations with an approximate kinematics. At the lowest bin, $Q^2=2 \; {\rm GeV^2}$, the differences seem to be smaller. This is due to the fact that in this region the calculation is dominated by the contribution of the quarks and the non-perturbative input which is the same in the $k_T$ and the collinear factorization formulae in our approach. The range in $x$ has been extended down to  $x=10^{-6}$ to cover the LHeC kinematic region.

In the lowest $Q^2$ region the quark contribution and the non-perturbative gluonic component becomes dominant. This is why the 
differences between the exact and the approximate kinematics are starting to become smaller in this  region as the whole $F_L$ gets larger contribution from the quarks which are treated in the same way in both of these calculations.

In the larger $x$ region, below $10^{-3}$, 
we observe that the kinematical effects are more significant.
Interestingly, the gluon contribution originating from $k_T>k_0$ at $x\simeq 0.01$ is very small for the scales $Q^2$ up to about $10~{\rm GeV^2}$ and this region is completely dominated by the quark  and the non-perturbative contribution. Therefore, using the approximate kinematics leads to  large overestimation of the perturbative gluon contribution in this region.

\section{Conclusions}
\label{sec:conclusions}

In this paper we have computed the longitudinal structure function $F_L$ of the proton within the $k_T$ factorization framework, using the unified BFKL/DGLAP resummation scheme for the unintegrated gluon density. Since we have only fitted our parameters to the $F_2$ data, the calculations for $F_L$ are absolute predictions. The calculations are consistent with the experimental data from HERA collider. 

We have analyzed the impact of the exact kinematics in the $k_T$ factorization scheme.  The exact gluon kinematics is very important for the phenomenological description of the data on $F_L$. In particular, it leads to larger differences than changing from the $k_T$ factorization to the collinear factorization scheme, at least in the available kinematical region of the HERA data.

We have also shown that the $k_T$ factorization scheme with exact kinematics includes both the collinear and the dipole limits. The first one is recovered by assuming the strong ordering in the gluon and quark momenta, together with the exact gluon kinematics. The latter one is recovered in the limit when the fraction of the gluon's longitudinal momentum is set to be equal to Bjorken $x$. 

The precision of the available HERA data for $F_L$ 
does not allow to uniquely discriminate between different approaches. 
It is possible however that by lowering the value of $Q^2$ or extending the kinematic regime to lower values of $x$ (like in the proposed LHeC collider) one can explore the differences between the presented frameworks.


\section*{Acknowledgements}
This work is partially supported by the grant MNiSW no. N202 249235. A.M.S. gratefully acknowledges the support  of the Alfred P. Sloan Research Foundation.

\bibliographystyle{h-physrev4}
\bibliography{mybib}



\end{document}